\documentclass[twocolumn,aps,pra,superscriptaddress,floatfix]{revtex4}
\usepackage{graphicx}
\usepackage{siunitx}
\usepackage{dcolumn}
\usepackage{bm}
\usepackage{kantlipsum}
\PassOptionsToPackage{hyphens}{url}\usepackage{hyperref}
\usepackage[mathlines]{lineno}
\usepackage{cleveref}
\usepackage{physics}
\usepackage{float}
\begin{document}
\title{Quantum-Enhanced continuous-wave stimulated Raman spectroscopy}

\author{R.\,B.\,Andrade}
\affiliation{Center for Macroscopic Quantum States bigQ, Department of Physics, Technical University of Denmark, Fysikvej 307, DK-2800 Kgs.\ Lyngby.}
\author{H.\,Kerdoncuff}
\affiliation{Danish Fundamental Metrology, Kogle Alle 5, DK-2970, Denmark.}
\author{K.\,Berg-Sørensen}
\affiliation{Center for Macroscopic Quantum States bigQ, Department of Physics, Technical University of Denmark, Fysikvej 307, DK-2800 Kgs.\ Lyngby.}
\author{T.\,Gehring}
\affiliation{Center for Macroscopic Quantum States bigQ, Department of Physics, Technical University of Denmark, Fysikvej 307, DK-2800 Kgs.\ Lyngby.}
\author{M.\,Lassen}
\affiliation{Danish Fundamental Metrology, Kogle Alle 5, DK-2970, Denmark.}
\author{U.\,L.\,Andersen}
\affiliation{Center for Macroscopic Quantum States bigQ, Department of Physics, Technical University of Denmark, Fysikvej 307, DK-2800 Kgs.\ Lyngby.}
\email{rabda@fysik.dtu.dk}

\begin{abstract}
Stimulated Raman spectroscopy has become a powerful tool to study the spatiodynamics of molecular bonds with high sensitivity, resolution and speed. However, sensitivity and speed of state-of-the-art stimulated Raman spectroscopy are currently limited by the shot-noise of the light beam probing the Raman process. Here, we demonstrate an enhancement of the sensitivity of continuous-wave stimulated Raman spectroscopy by reducing the quantum noise of the probing light below the shot-noise limit by means of amplitude squeezed states of light. Probing polymer samples with Raman shifts around \SI{2950}{\centi\meter}$^{-1}$ with squeezed states, we demonstrate a quantum-enhancement of the stimulated Raman signal-to-noise ratio (SNR) of \SI{3.60}{\decibel} relative to the shot-noise limited SNR. 
Our proof-of-concept demonstration of quantum-enhanced Raman spectroscopy paves the way for a new generation of Raman microscopes, where weak Raman transitions can be imaged without the use of markers or an increase in the total optical power.
\end{abstract}
\maketitle

\section{Introduction}

Optical quantum sensing exploits the unique quantum correlations of non-classical light to enhance the detection of physical parameters beyond classical means~\cite{Lawrie2019QuantumLight, Pirandola2018AdvancesSensing, Marshall2016Continuous-variableData, Taylor2016QuantumBiology, Hoff2013Quantum-enhancedSensitivity}. While several different quantum states of light can, in principle, be used to provide such a quantum advantage, so far, it is only the ubiquitous squeezed states of light that have demonstrably shown to provide a real practical advantage~\cite{Lvovsky2015SqueezedLight, Andersen201630Generation, Schnabel2017}. Squeezed states of light have for example enabled quantum-enhanced measurements of mechanical displacements~\cite{Hoff2013Quantum-enhancedSensitivity,Pooser:15}, magnetic fields~\cite{Mitchell2019Sqzopticalmagnetometry, Li:18}, viscous-elasticity of cells~\cite{Taylor2013BiologicalLimit} and, most prominently, gravitational waves~\cite{LIGO2013}. Another field that could significantly benefit from quantum-enhanced sensing by means of squeezed light -- but yet not demonstrated -- is Stimulated Raman Spectroscopy (SRS).\par   
SRS is a very powerful technique to perform real-time vibrational imaging of living cells and organisms and it has therefore provided a deeper understanding of properties of biological systems~\cite{Freudiger2014StimulatedSource, Camp2015ChemicallyScattering,Cheng2016CoherentMicroscopy, Jones2019RamanFrontiers}. It is based on the stimulated excitation of a Raman transition of the sample under interrogation, thereby resulting in a measurable stimulated Raman loss and gain of the two input beams, respectively. It allows for non-invasive and in-vivo measurements with short acquisition times~\cite{Lee2017ImagingMicroscopy} and has enabled the structural and dynamical imaging of lipids~\cite{Dou2012Label-freeMicroscopy, Wang2011RNAiMicroscopy} as well as the characterization of healthy and tumorous brain tissues~\cite{Ji2013RapidMicroscopy,Lu2016Label-freeImaging}.\par
In SRS, the sensitivity and the imaging speed are fundamentally limited by the noise level (often shot-noise) of the probing laser~\cite{Moester2015OptimizedMicroscopy,Kerdoncuff2019Continuous-waveShifts} but can in principle be arbitrarily improved simply by increasing the power of the input beams. However, in biological system, especially in living systems, the power must be kept low to avoid changing the biological dynamics of the specimens or even damaging it due to excessive heating. Leaving the optical power at a constant level, the sensitivity and bandwidth of the SRS can be boosted by reducing the shot-noise level using squeezed states of light. \par
In this article, we demonstrate the quantum enhancement of continuous-wave (cw) SRS using amplitude squeezed light. We demonstrate its functionality and superiority by spectroscopically measuring the carbon-hydrogen (C-H) vibrations of polymethylmethacrylate (PMMA) and polydimethylsiloxane (PDMS) with a sensitivity-improvement of approximately 56\% relative to shot-noise limited Raman spectrometer. Our measurement method has the potential to enable new measurement regimes of Raman bio-imaging that are inaccessible by the conventional shot-noise limited Raman spectrometer. \par

\section{Basic concept}

SRS employs two laser beams -- known as the pump and the probe (Stokes) laser beams -- to coherently excite a selected molecular vibration of the system under investigation. If the vibrational frequency of the chemical bond matches the frequency difference of the pump and probe laser, the Raman interaction is stimulated and as a result significantly amplified by orders of magnitude. In the stimulated Raman effect, a photon is annihilated from the pump beam and simultaneously a Raman shifted photon is created in the background noise of the probe beam. This background noise is fundamentally limited by shot-noise when using a probe beam in a coherent state from a conventional laser, but it can be reduced by using a laser beam in a squeezed state. By using a bright amplitude squeezed beam (squeezed state with a large coherent excitation), the quantum-enhanced sensitivity is directly proportional to the standard deviation of the squeezed amplitude quadrature, $\Delta X$: $\delta \propto \Delta X/
\sqrt{I_{SRS}}$. Here $I_{SRS}$ is the intensity of the SRS signal which changes linearly with the intensity of the probe ($I_{s}$) and pump ($I_{p}$) as $I_{SRS}=K I_{p} I_{s}$, where $K$ is a constant related to the number of probed molecules and their Raman cross section~\cite{Boyd2008NonlinearOptics}. It is thus clear that the sensitivity can be improved without changing the power by reducing the noise of the amplitude quadrature.

\section{Experimental setup}\label{ExpSetup}
\begin{figure*}
	\begin{center}
		\includegraphics[width =\textwidth]{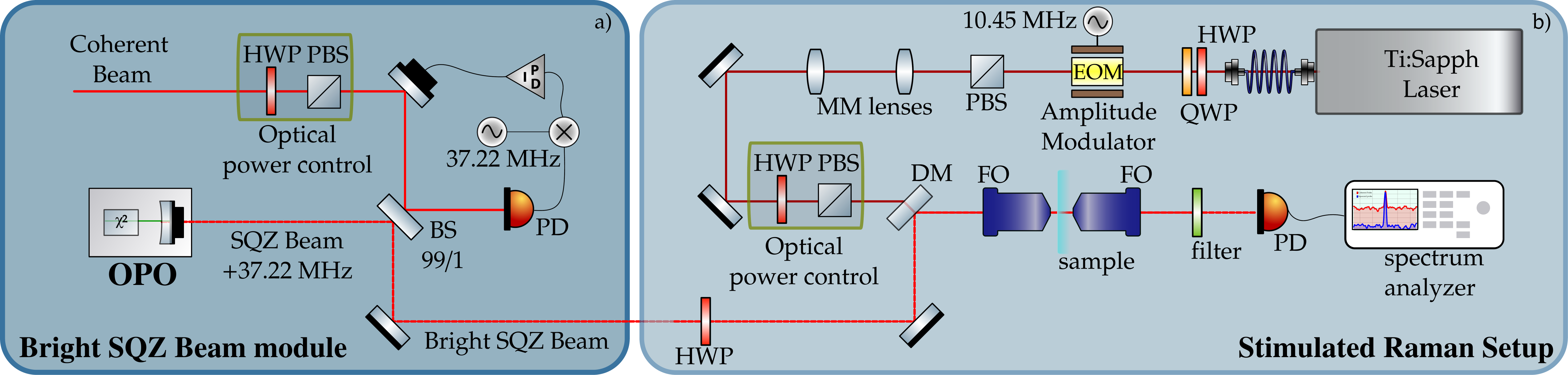}
		\caption{Experimental setup to measure the SRS signal enhanced by squeezed light. a) Bright squeezed beam preparation at \SI{1064}{\nano\meter}. The squeezed state is combined with the coherent beam in a 99/1 beam splitter (BS) to generate a bright squeezed state at one output serving as the probe for the SRS setup while the other output is used to phase lock the relative phase of the two beams (using a phase modulation at \SI{37.22}{\mega\hertz}). b) SRS setup. A wavelength-tunable Ti:Sapph laser is intensity modulated at 
the frequency of \SI{10.45}{\mega\hertz} and serve as the pump beam. The pump and probe beams are overlapped in a dichroic mirror (DM) and focused into the sample using a microscope objective (FO). The SRS signal is collected using a second objective, the pump beam is filtered off and the probe beam is measured using a high quantum efficiency (HQE) photodiode. The results are acquired by an electronic spectrum analyzer (ESA) by which a power spectrum of the signal is attained. HWP: half wave plate, PBS: polarizing beam splitter, PD: photodiode, BS: beam splitter, QWP: quarter wave plate, DM: dichroic mirror, FO: focal objective, EOM: electro-optical modulator , MM lenses: mode matching lenses.}
 		\label{fig_setup}
 		\end{center}
\end{figure*}

The experimental setup is shown in Fig.~\ref{fig_setup}. It consists of two modules: the bright squeezed light module and the SRS module as will now be discussed in detail.\par
{\it Bright Squeezed Light Module:} The laser source was an Innolight GmbH Diabolo operating at \SI{1064}{\nano\meter} with an internal module for second harmonic generation (SHG) at \SI{532}{\nano\meter}. The squeezed state was generated in a linear optical parametric oscillator (OPO) cavity consisting of a periodically poled potassium titanyl phosphate (PPKTP) crystal and a hemispheric coupling mirror. When pumping with a power of \SI{80}{\milli\watt} at \SI{532}{\nano\meter}, setting the phase of the pump beam to deamplification and injecting a seed beam with a power of \SI{600}{\micro\watt} at \SI{1064}{\nano\meter}, the OPO produced \SI{7}{\decibel} of amplitude squeezed light. More details about the squeezed light source can be found in~\cite{Schafermeier2018DeterministicSuper-resolution}. The amplitude squeezed light and a coherent beam at \SI{1064}{\nano\meter} were combined on an asymmetric (99/1) beam splitter to produce a bright amplitude squeezed beam. The phase between these beams was actively stabilized by feeding a phase shifter in the coherent beam path with an error signal that was generated by electronically demodulating the photo detected beat of the bright coherent beam and the \SI{37.22}{\mega\hertz} phase modulation side-bands accompanying the squeezed field. The output of the $99\,\%$ port of the BS was sent to the SRS module serving as the probe beam for Raman spectroscopy.\par
{\it Stimulated Raman Module:} The pump beam for SRS was a tunable Ti:Sapph laser (MSquare SolsTiS) scanned from \SI{800}{\nano\meter} to \SI{830}{\nano\meter}. It delivered a maximum output power of \SI{200}{\milli\watt} which could be adjusted at the entrance to the microscope. The pump beam intensity was modulated at \SI{10.45}{\mega\hertz} with a sinusoidal function using a resonant electro-optical amplitude modulator. The beam size of the pump beam was adjusted with a set of lenses (MM lenses) in order to optimize the overlap with the probe beam. A fine adjustment in the polarization between the pump and the probe beams was made using a HWP (half-wave plate) in the probe path. After combining the probe and the pump beams at a dichroic mirror, both beams were focused to a spot size of \SI{2.5}{\micro\meter} on the sample with a 20x microscope objective. The beams were collected and collimated by a second microscope objective after which the pump beam was filtered using a long-pass filter and the probe beam was detected using a photodiode with a quantum efficiency of more than $ 99\,\%$ (Fermionics InGaAs FD500). The stimulated Raman gain was deduced from the power spectrum which was recorded using an electrical spectrum analyzer.\par
An important factor when using squeezed light are the optical losses in the optical pathway of the squeezed beam. From the output of the OPO cavity to the entrance of the microscope we estimated an overall optical efficiency of around $\eta_{path}$ = $85\,\%$, while each of the two microscope objectives had a transmission efficiency of $97\,\%$. The visibility between the coherent and the squeezed beam was $95\,\%$. Thus, the total efficiency transmission of the \SI{1064}{\nano\meter} path including also the detection losses was estimated to $\eta_{total}$=$67\,\%$. \par
\begin{figure}[htpb]
	\begin{center}
		\includegraphics[width =0.45\textwidth]{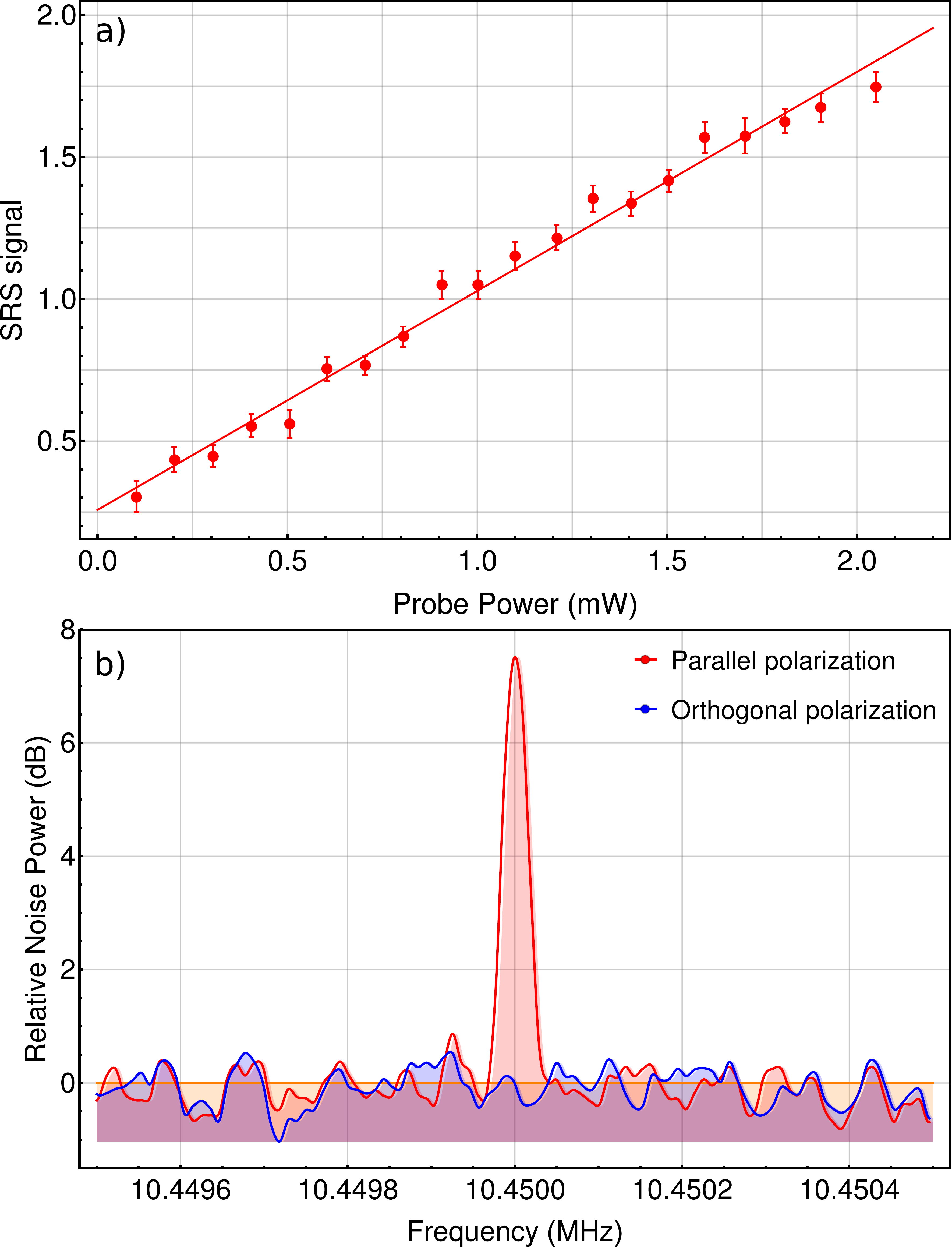}
		\caption{Characterization of the SRS signal using an average pump power of \SI{38}{\milli\watt}. a) Intensity of the SRS signal as a function of the probe power. b) Polarization behavior of the SRS signal using \SI{1.35}{\milli\watt} of power in the probe beam. The red trace represents the signal when the probe and pump beams are parallel polarized while the blue trace is associated to orthogonal polarized beams. All traces are normalized to the shot-noise level.}
 		\label{fig_SRSclass}
 		\end{center}
\end{figure}

In this work we use two different solid samples to characterize the SRS process, PMMA and PDMS. Both samples have Raman transitions in the region between \SIrange[range-units=single,range-phrase=--]{2800}{3100}{\centi\meter}$^{-1}$ corresponding to vibration modes of C-H bonds~\cite{Dirlikov1980Carbonhydrogenmethylmethacrylate, Smith1984VIBRATIONALANALOGS.}. We start by classically characterizing the Raman transition of a PMMA sample of \SI{2}{\milli\meter} thickness and a pump laser with a power at the sample of \SI{38}{\milli\watt} and tuned to the wavelength of \SI{810.241}{\nano\meter} to hit the Raman transition at \SI{2948.32}{\centi\meter}$^{-1}$. The SRS signal was measured on the probe beam (due to the stimulated gain) at the modulation frequency of the pump at \SI{10.45}{\mega\hertz}, and we acquired a power spectrum around this frequency. In absence of the SRS signal only measurement noise was detected. The data presented have all been measured using a resolution bandwidth of \SI{30}{\hertz} and a video bandwidth of \SI{1}{\hertz}, each data point was averaged 30 times and the electronic noise was subtracted in all the measurements. The probe power was changed from \SI{250}{\micro\watt} to \SI{2.0}{\milli\watt}, as shown in Fig.~\ref{fig_SRSclass}a) and we clearly observe the expected linear dependency between SRS signal and probe power. The polarization behavior between pump and probe beams are shown in Fig.~\ref{fig_SRSclass}b) where the red trace represents the signal when the pump and probe beams were parallel polarized while the blue trace corresponds to the signal when the beams were orthogonal polarized. It is clear that the Raman signal disappears in the latter case, thus further corroborating the presence of real Raman signal in the former  case~\cite{Tanaka2006ReviewPolymers,Kerdoncuff2017CompactSpectroscopy}. Both traces were normalized by the shot-noise.\par    
Having verified the C-H Raman transition, in the following we present the demonstration of quantum-enhanced SRS. To clearly demonstrate quantum-improved performance beyond the conventional approach, we conducted the experiment both with the probe beam in a coherent state (limited by shot-noise and representing the conventional approach) and in the squeezed state. The experimental scheme could easily be swapped between the two modes of operation simply by blocking and unblocking the squeezed vacuum state which will have no effect on the probe or pump input powers. 
\section{Experimental Results}
\begin{figure}[htpb]
	\begin{center}
		\includegraphics[width =0.45\textwidth]{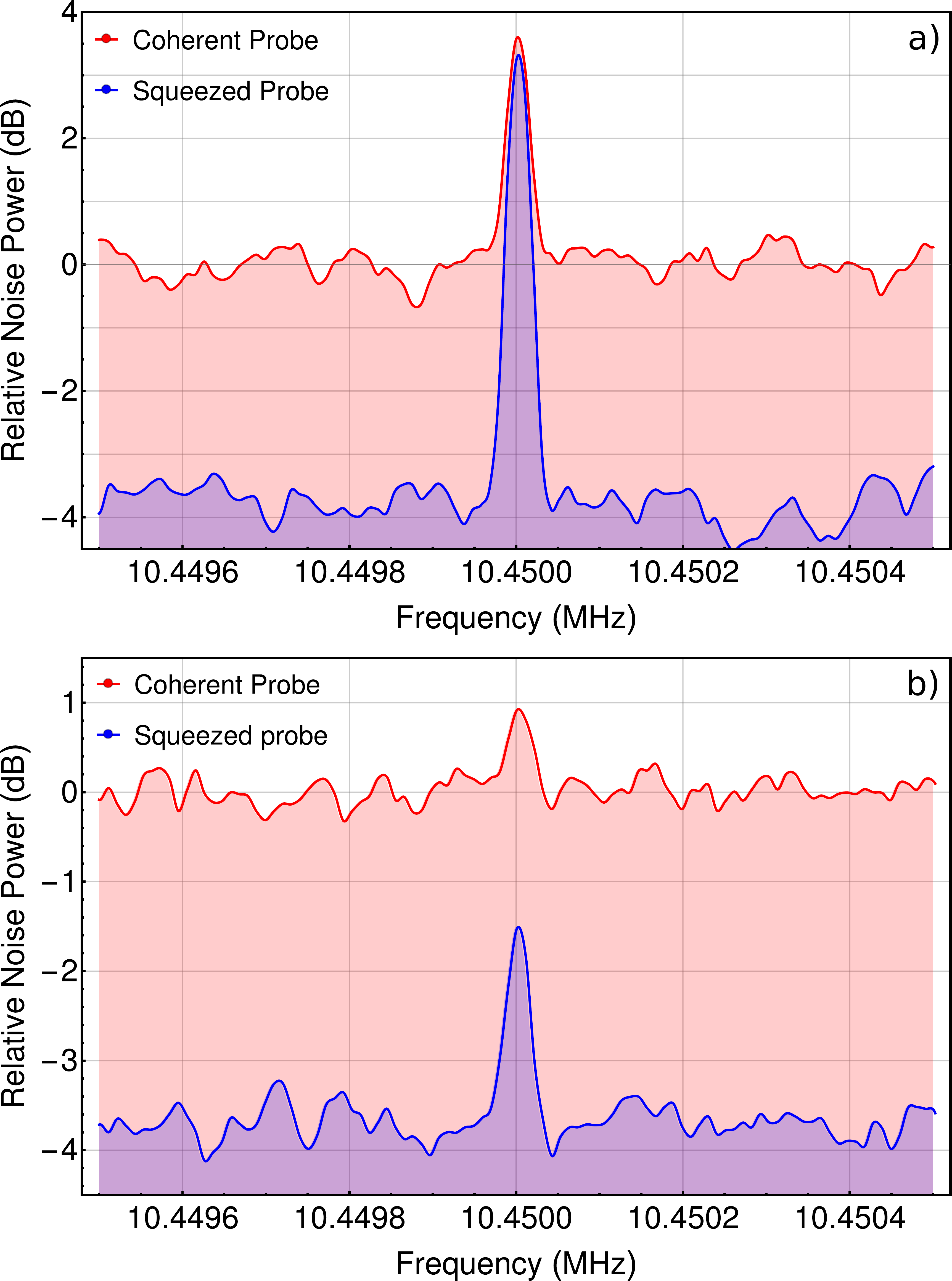}
		\caption{Demonstration of quantum enhanced SRS using probe powers of \SI{1.3}{\milli\watt} and pump powers of: a) \SI{24}{\milli\watt} and b) \SI{11}{\milli\watt}. The red SRS traces correspond to the realizations where the probe beams are in a coherent state while the blue traces correspond to the beams being in a squeezed state with -\SI{3.60}{\decibel} noise suppression below the shot-noise. In both cases the signals are normalized to the shot-noise level.}
 		\label{fig_PMMA_power}
 		\end{center}
\end{figure}

Figure~\ref{fig_PMMA_power} presents our experimental results for quantum-enhanced SRS. We present the spectra for the Raman shift of PMMA using both a coherent state (for comparison) and a squeezed state with optical powers of \SI{1.3}{\milli\watt} while the pump power was set to~\SI{24}{\milli\watt} (Fig.~\ref{fig_PMMA_power}a) and~\SI{11}{\milli\watt}(Fig.~\ref{fig_PMMA_power}b). It is clear from the spectra that the usage of squeezed light significantly improves the signal-to-noise ratio and therefore the sensitivity of the Raman spectrometer. We see in particular that for pump powers lower than around~\SI{11}{\milli\watt}, the Raman signal is almost embedded in shot-noise and only becomes pronounced when using squeezed states of light. It is therefore clear that by using the quantum-enhanced operation mode, it is possible to attain Raman signals even for low pump powers. This is of importance when studying fragile biological systems where excessive powers might change the dynamics of the system.\par
\begin{figure}[htpb]
	\begin{center}
		\includegraphics[scale=0.24]{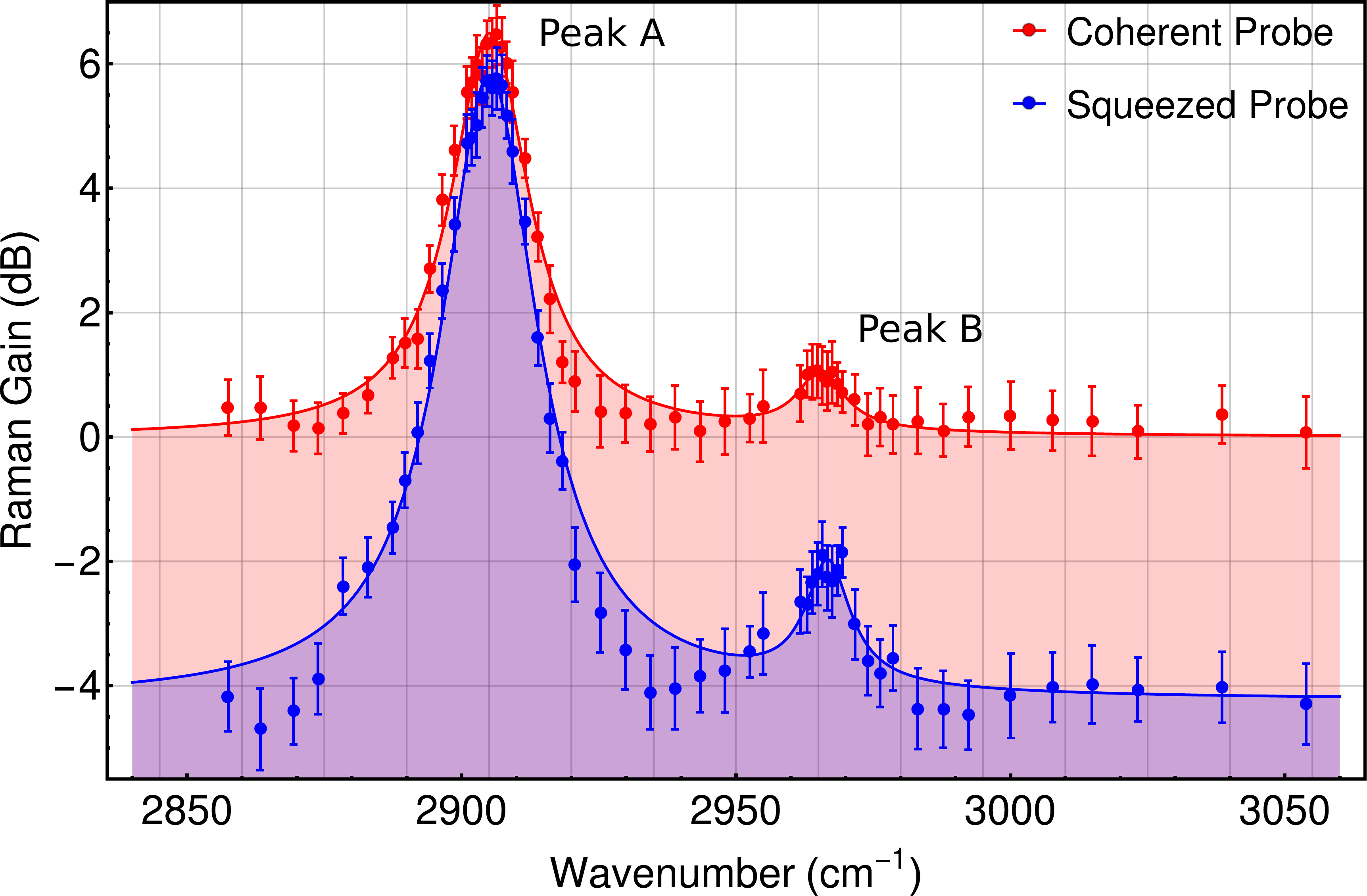}
		\caption{SRS spectrum of PDMS. The pump beam is scanned around the C-H stretching region with pump and probe powers of \SI{28}{\milli\watt} and \SI{1.3}{\milli\watt}, respectively. The traces are normalized to the shot-noise level.}
 		\label{fig_PDMS_spec}
 		\end{center}
\end{figure}

The SRS process provides a Raman spectrum similar to the spectrum generated using Spontaneous Raman Spectroscopy techniques. Using a PDMS sample and sweeping the pump laser, manually, from \SI{803.36}{\nano\meter} to \SI{816.36}{\nano\meter}, the Raman spectrum of C-H stretching modes in the region between \SIrange[range-units=single,range-phrase=-]{2850}{3100}{\centi\meter}$^{-1}$ was acquired and it is depicted in Fig.~\ref{fig_PDMS_spec}. The probe and the pump optical powers were \SI{1.3}{\milli\watt} and \SI{28}{\milli\watt}, respectively. While scanning the wavelength of the pump laser, the optical pump power was continuously measured and used to normalize the acquired Raman spectrum at every wavelength. In Fig.~\ref{fig_PDMS_spec} the spectra are shown for coherent (red trace) and squeezed states (blue trace). Lorentzian multi-peak fits were used to obtain the two Raman shifts in table \ref{table_PDMS}.\par
\begin{figure*}
	\begin{center}
		\includegraphics[width =0.8\textwidth]{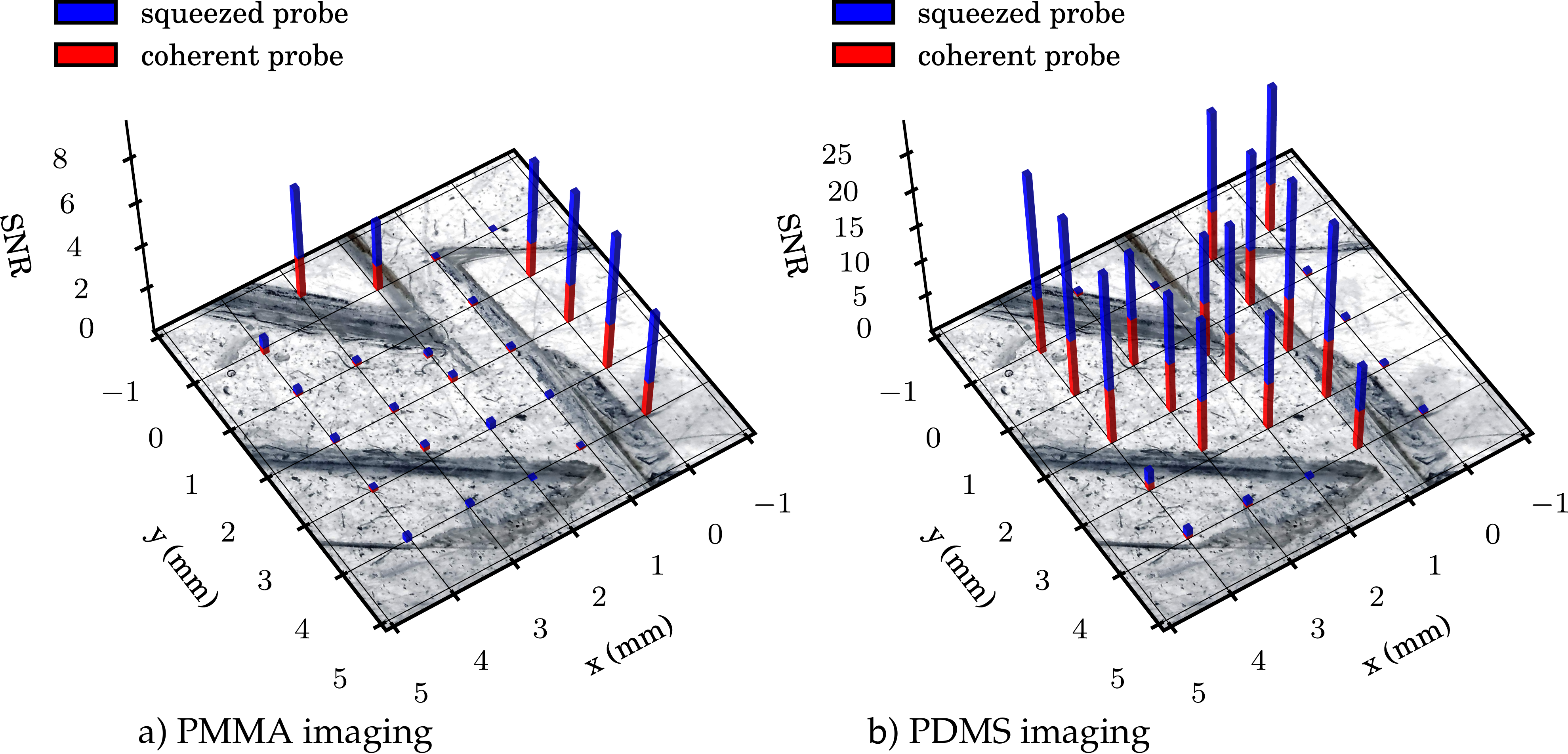}
		\caption{SRS imaging of different polymers in a sample comprising PDMS, PMMA and polystyrene. a) Imaging of SRS resonances attained when the pump laser frequency was set to reach the PMMA vibrational mode  \SI{2948.75}{\centi\meter}$^{-1}$. b) Imaging the vibrational mode \SI{2904.76}{\centi\meter}$^{-1}$ of PDMS. The remaining area (polystyrene) does not produce an SRS signal.}
 		\label{fig_Imaging}
 		\end{center}
\end{figure*}

\begin{table*}
\begin{ruledtabular}
\begin{tabular}{llcccc}
Probe Beam & Assignment & \multicolumn{1}{c}{\textrm{$ \nu_{Peak A}$ (\SI{}{\centi\meter}$^{-1}$)}} &
\multicolumn{1}{c}{\textrm{$ SNR_{A} (dB)$}}&
\multicolumn{1}{c}{\textrm{$\nu_{Peak B}$ (\SI{}{\centi\meter}$^{-1}$)}}&
\multicolumn{1}{c}{\textrm{$ SNR_{B} (dB)$}}\\
\hline
Coherent light & C-H sym & \mbox{2905.17 (0.15)} &\mbox{6.49 (0.08)} &\mbox{2965.46 (1.17)} &\mbox{0.95 (0.18)}\\
Ampl. squeezed light & C-H sym & \mbox{2905.07 (0.14)} &\mbox{10.05 (0.07)} &\mbox{2966.74 (0.80)} &\mbox{2.01 (0.18)}\\
\end{tabular}
\end{ruledtabular}
\caption{\label{table_PDMS} Raman resonances of PDMS: Peak A and Peak B. The Raman shifts are represented as $\nu_{Peak A}$ and $\nu_{Peak B} $.}
\end{table*}
To demonstrate the potential of using the quantum-enhanced Raman spectrometer as a microscope, we performed a rough raster scan of a sample consisting of three different polymers; PMMA, PDMS and polystyrene. A 3-axes translational stage with differential micrometer screws was used to move, manually, the sample position in steps of \SI{1}{\milli\meter} in a square region of 7x7 mm$^2$. The SRS signal was acquired using coherent and squeezed states of light alternately for each displacement. Applying an average pump power of \SI{28}{\milli\watt} and a probe power of \SI{1.3}{\milli\watt}, the pump laser wavelength is set up to \SI{810.213}{\nano\meter} corresponding to a Raman shift \SI{2948.75}{\centi\meter}$^{-1}$ and the PMMA content in the sample was detected. Fig.~\ref{fig_Imaging} a) shows the result. Afterwards, to detect the PDMS content in the sample, the pump wavelength was changed to \SI{813.111}{\nano\meter} corresponding to the vibrational mode \SI{2904.76}{\centi\meter}$^{-1}$. The result is shown in Fig.~\ref{fig_Imaging} b). The remaining area comprising polystyrene exhibits no signals as it has no vibrational modes in the interrogated frequency region.
We clearly see from the figure that PMMA and PDMS can be  distinguished with the method, and we also find that squeezed light outperforms coherent light operation in the entire imaging plane. These proof-of-concept imaging measurements demonstrate the usefulness of the quantum-enhanced Raman method for microscopy.\par
\section{Conclusion}
In summary, we demonstrated a sensitivity enhancement of the stimulated Raman process using squeezed states of light. The quantum enhancement was measured to be more than 50\% in comparison to the conventional approach with coherent states. Our technique was used to visualize spectroscopically the Raman bands within the C-H stretching region of polymer samples (PMMA and PDMS) and to perform chemically specific imaging measurements. The sensitivity of our quantum spectrometer can be further improved by minimizing the optical losses of the system and by employing states with a higher degrees of squeezing. Moreover, to realize real and high-resolution SRS imaging, the sample should be scanned with high spatial resolution and the objectives replaced with ones with higher numerical apertures.\par
We believe that our demonstration opens the door to new possibilities for Raman spectroscopy and microscopy. Using squeezed light to enhance the sensitivity of the stimulated Raman signal enables studies of biological samples with a lower risk of damage due to high beam powers. This might enable the studies of bio-physical effects that may not be visible using the standard classical approaches. The presented method is not limited to the wavenumber range investigated in this work but can be extended to the fingerprint region (\SIrange[range-units=single,range-phrase=--]{500}{1800}{\centi\meter}$^{-1}$) by appropriate choice of laser wavelengths, thereby giving access to detailed information and the rich dynamics of different biological samples.
\section*{Funding}
The work was partly supported by the Danish National Research Foundation through the Center for Macroscopic Quantum States (bigQ, DNRF412), partly by the Innovation Foundation Denmark (Qubiz). Hugo Kerdoncuff and Mikael Lassen acknowledge financial support from the Danish Agency for Institutions and Educational Grants and EMPIR project 17FUN01-BeCOMe. The EMPIR initiative is funded by the European Union Horizon 2020 research and innovation program and co-financed by the EMPIR participating states.
\section*{Disclosures}
The authors declare no conflicts of interest.
\bibliographystyle{unsrt}
\bibliography{literature}

\begin{thebibliography}{10}

\bibitem{Lawrie2019QuantumLight}
B.~J. Lawrie, P.~D. Lett, A.~M. Marino, and R.~C. Pooser.
\newblock {Quantum Sensing with Squeezed Light}.
\newblock {\em ACS Photonics}, 6(6):1307--1318, 6 2019.

\bibitem{Pirandola2018AdvancesSensing}
S.~Pirandola, B.~R. Bardhan, T.~Gehring, C.~Weedbrook, and S.~Lloyd.
\newblock {Advances in photonic quantum sensing}.
\newblock {\em Nature Photonics}, 12(12):724--733, 2018.

\bibitem{Marshall2016Continuous-variableData}
K.~Marshall, C.~S. Jacobsen, C.~Sch{\"{a}}fermeier, T.~Gehring, C.~Weedbrook,
  and U.~L. Andersen.
\newblock {Continuous-variable quantum computing on encrypted data}.
\newblock {\em Nature Communications}, 7, 12 2016.

\bibitem{Taylor2016QuantumBiology}
Michael~A. Taylor and Warwick~P. Bowen.
\newblock {Quantum metrology and its application in biology}.
\newblock {\em Physics Reports}, 615:1--59, 2 2016.

\bibitem{Hoff2013Quantum-enhancedSensitivity}
Ulrich~B. Hoff, Glen~I. Harris, Lars~S. Madsen, Hugo Kerdoncuff, Mikael Lassen,
  Bo~M. Nielsen, Warwick~P. Bowen, and Ulrik~L. Andersen.
\newblock {Quantum-enhanced micromechanical displacement sensitivity}.
\newblock {\em Optics Letters}, 38(9):1413, 5 2013.

\bibitem{Lvovsky2015SqueezedLight}
A.~I. Lvovsky.
\newblock {Squeezed Light}.
\newblock In David~L. Andrews, editor, {\em Photonics}, chapter~5, pages
  121--163. John Wiley and Sons, Ltd, 2015.

\bibitem{Andersen201630Generation}
Ulrik~L Andersen, Tobias Gehring, Christoph Marquardt, and Gerd Leuchs.
\newblock 30 years of squeezed light generation.
\newblock {\em Physica Scripta}, 91(5):053001, 4 2016.

\bibitem{Schnabel2017}
Roman Schnabel.
\newblock Squeezed states of light and their applications in laser
  interferometers.
\newblock {\em Physics Reports}, 684, 11 2016.

\bibitem{Pooser:15}
Raphael~C. Pooser and Benjamin Lawrie.
\newblock Ultrasensitive measurement of microcantilever displacement below the
  shot-noise limit.
\newblock {\em Optica}, 2(5):393--399, 5 2015.

\bibitem{Mitchell2019Sqzopticalmagnetometry}
Florian Wolfgramm, Alessandro Cer\`e, Federica~A. Beduini, Ana
  Predojevi\ifmmode~\acute{c}\else \'{c}\fi{}, Marco Koschorreck, and Morgan~W.
  Mitchell.
\newblock Squeezed-light optical magnetometry.
\newblock {\em Phys. Rev. Lett.}, 105:053601, 7 2010.

\bibitem{Li:18}
Bei-Bei Li, Jan B\'{i}lek, Ulrich~B. Hoff, Lars~S. Madsen, Stefan Forstner,
  Varun Prakash, Clemens Sch\"{a}fermeier, Tobias Gehring, Warwick~P. Bowen,
  and Ulrik~L. Andersen.
\newblock Quantum enhanced optomechanical magnetometry.
\newblock {\em Optica}, 5(7):850--856, Jul 2018.

\bibitem{Taylor2013BiologicalLimit}
Michael~A. Taylor, Jiri Janousek, Vincent Daria, Joachim Knittel, Boris Hage,
  Hans~A. Bachor, and Warwick~P. Bowen.
\newblock {Biological measurement beyond the quantum limit}.
\newblock {\em Nature Photonics}, 7(3):229--233, 3 2013.

\bibitem{LIGO2013}
The LIGO~Scientific Collaboration.
\newblock Enhanced sensitivity of the ligo gravitational wave detector by using
  squeezed states of light.
\newblock {\em Nature Photonics}, 7(8):613 -- 619, 2013.

\bibitem{Freudiger2014StimulatedSource}
Christian~W. Freudiger, Wenlong Yang, Gary~R. Holtom, Nasser Peyghambarian,
  X.~Sunney Xie, and Khanh~Q. Kieu.
\newblock {Stimulated Raman scattering microscopy with a robust fibre laser
  source}.
\newblock {\em Nature Photonics}, 8(2):153--159, 2 2014.

\bibitem{Camp2015ChemicallyScattering}
Charles~H. Camp and Marcus~T. Cicerone.
\newblock {Chemically sensitive bioimaging with coherent Raman scattering}.
\newblock {\em Nature Photonics}, 9(5):295--305, 5 2015.

\bibitem{Cheng2016CoherentMicroscopy}
J.~X. Cheng and X.~S. Xie.
\newblock {\em {Coherent Raman scattering microscopy}}.
\newblock CRC Press, 1st edition edition, 1 2016.

\bibitem{Jones2019RamanFrontiers}
Robin~R. Jones, David~C. Hooper, Liwu Zhang, Daniel Wolverson, and
  Ventsislav~K. Valev.
\newblock {Raman Techniques: Fundamentals and Frontiers}.
\newblock {\em Nanoscale Research Letters}, 14(1), 12 2019.

\bibitem{Lee2017ImagingMicroscopy}
Hyeon~Jeong Lee and Ji-Xin Cheng.
\newblock Imaging chemistry inside living cells by stimulated raman scattering
  microscopy.
\newblock {\em Methods}, 128:119 -- 128, 2017.

\bibitem{Dou2012Label-freeMicroscopy}
Wei Dou, Delong Zhang, Yookyung Jung, Ji~Xin Cheng, and David~M. Umulis.
\newblock {Label-free imaging of lipid-droplet intracellular motion in early
  Drosophila embryos using femtosecond-stimulated Raman loss microscopy}.
\newblock {\em Biophysical Journal}, 102(7):1666--1675, 4 2012.

\bibitem{Wang2011RNAiMicroscopy}
Meng~C. Wang, Wei Min, Christian~W. Freudiger, Gary Ruvkun, and X.~Sunney Xie.
\newblock {RNAi screening for fat regulatory genes with SRS microscopy}.
\newblock {\em Nature Methods}, 8(2):135--138, 2 2011.

\bibitem{Ji2013RapidMicroscopy}
Minbiao Ji, Daniel~A. Orringer, Christian~W. Freudiger, Shakti Ramkissoon,
  Xiaohui Liu, Darryl Lau, Alexandra~J. Golby, Isaiah Norton, Marika Hayashi,
  Nathalie~Y.R. Agar, Geoffrey~S. Young, Cathie Spino, Sandro Santagata, Sandra
  Camelo-Piragua, Keith~L. Ligon, Oren Sagher, and X.~Sunney~Xie.
\newblock {Rapid, label-free detection of brain tumors with stimulated raman
  scattering microscopy}.
\newblock {\em Science Translational Medicine}, 5(201), 9 2013.

\bibitem{Lu2016Label-freeImaging}
Fa~Ke Lu, David Calligaris, Olutayo~I. Olubiyi, Isaiah Norton, Wenlong Yang,
  Sandro Santagata, X.~Sunney Xie, Alexandra~J. Golby, and Nathalie~Y.R. Agar.
\newblock {Label-free neurosurgical pathology with stimulated Raman imaging}.
\newblock {\em Cancer Research}, 76(12):3451--3462, 6 2016.

\bibitem{Moester2015OptimizedMicroscopy}
M.~J. Moester, F.~Ariese, and J.~F. De~Boer.
\newblock {Optimized signal-to-noise ratio with shot noise limited detection in
  stimulated raman scattering microscopy}.
\newblock {\em Journal of the European Optical Society}, 10, 2015.

\bibitem{Kerdoncuff2019Continuous-waveShifts}
Hugo Kerdoncuff, Mikael Lassen, and Jan~C. Petersen.
\newblock {Continuous-wave coherent Raman spectroscopy for improving the
  accuracy of Raman shifts}.
\newblock {\em Optics Letters}, 44(20):5057, 10 2019.

\bibitem{Boyd2008NonlinearOptics}
Robert~W. Boyd.
\newblock {\em {Nonlinear Optics}}.
\newblock Academic Press, 2008.

\bibitem{Schafermeier2018DeterministicSuper-resolution}
Clemens Sch{\"{a}}fermeier, Miroslav Je{\v{z}}ek, Lars~S. Madsen, Tobias
  Gehring, and Ulrik~L. Andersen.
\newblock {Deterministic phase measurements exhibiting super-sensitivity and
  super-resolution}.
\newblock {\em Optica}, 5(1):60, 1 2018.

\bibitem{Dirlikov1980Carbonhydrogenmethylmethacrylate}
Stoil Dirlikov and Jack~L. Koenig.
\newblock {Carbon–hydrogen stretching and bending vibrations in the Raman
  spectra of poly (methylmethacrylate)}.
\newblock {\em Journal of Raman Spectroscopy}, 9(3):150--154, 6 1980.

\bibitem{Smith1984VIBRATIONALANALOGS.}
A.~Lee Smith and Dennis~R. Anderson.
\newblock {VIBRATIONAL SPECTRA OF Me2SiCl2, Me3SiCl, Me3SiOSiMe3, (Me2SiO)3,
  (Me2SiO)4, (Me2SiO)x, AND THEIR DEUTERATED ANALOGS.}
\newblock {\em Applied Spectroscopy}, 38(6):822--834, 1984.

\bibitem{Tanaka2006ReviewPolymers}
M.~Tanaka and R.~J. Young.
\newblock {Review Polarised Raman spectroscopy for the study of molecular
  orientation distributions in polymers}.
\newblock {\em Journal of Materials Science}, 41(3):963--991, 2 2006.

\bibitem{Kerdoncuff2017CompactSpectroscopy}
Hugo Kerdoncuff, Mark~R. Pollard, Philip~G. Westergaard, Jan~C. Petersen, and
  Mikael Lassen.
\newblock {Compact and versatile laser system for polarization-sensitive
  stimulated Raman spectroscopy}.
\newblock {\em Optics Express}, 25(5):5618, 3 2017.

\end{thebibliography}
\end{document}